# A note on rate-distortion functions for nonstationary Gaussian autoregressive processes

Robert M. Gray and Takeshi Hashimoto *

June 29, 2018

**Abstract**

Source coding theorems and Shannon rate-distortion functions were studied for the discrete-time Wiener process by Berger and generalized to nonstationary Gaussian autoregressive processes by Gray and by Hashimoto and Arimoto. Hashimoto and Arimoto provided an example apparently contradicting the methods used in Gray, implied that Gray's rate-distortion evaluation was not correct in the nonstationary case, and derived a new formula that agreed with previous results for the stationary case and held in the nonstationary case. In this correspondence it is shown that the rate-distortion formulas of Gray and Hashimoto and Arimoto are in fact consistent and that the example of of Hashimoto and Arimoto does not form a counter example to the methods or results of the earlier paper. Their results do provide an alternative, but equivalent, formula for the rate-distortion function in the nonstationary case and they provide a concrete example that the classic Kolmogorov formula differs from the autoregressive formula when the autoregressive source is not stationary. Some observations are offered on the different versions of the Toeplitz asymptotic eigenvalue distribution theorem used in the two papers to emphasize how a slight modification of the classic theorem avoids the problems with certain singularities.

## 1 Introduction

A Gaussian autoregressive source is defined by the difference equation

$$X_n = \begin{cases} -\sum_{k=1}^{n} a_k X_{n-k} + Z_n & n = 1, 2, \cdots \\ 0 & n \leq 0 \end{cases} \tag{1}$$

where the $Z_n$ are iid random variables with mean zero and variance $\sigma^2$ and where we require that

$$\sum_{k=0}^{\infty} |a_k| < \infty \tag{2}$$

for $a_0 = 1$. If the roots of $A(z) = \sum_{k=0}^{\infty} a_k z^{-k}$ all lie strictly inside of unit circle, then the statistics of the process approach a stationary distribution and the Shannon rate-distortion

*Robert M. Gray is with the Information Systems Laboratory, Electrical Engineering Dept., Stanford University, Stanford, CA 94305. Takeshi Hashimoto is with the Dept. of Electronics Eng., University of Electro-Communications, 1-5-1 Chofugaoka, Chofushi, Tokyo 182-8585, Japan

function of the process is given parametrically by Kolmogorov's classic formula [4] (see also [5])

$$
\begin{aligned}
D_\theta &= \frac{1}{2\pi}\int_{-\pi}^{\pi} \min\left[\theta,\ \frac{\sigma^2}{g(\omega)}\right] d\omega && (3)\\
R(D_\theta) &= \frac{1}{2\pi}\int_{-\pi}^{\pi} \max\left[\frac{1}{2}\ln\frac{\sigma^2}{\theta g(\omega)},\ 0\right] d\omega && (4)\\
g(\omega) &= \left|\sum_{k=0}^{\infty} a_k e^{-jk\omega}\right| && (5)
\end{aligned}
$$

for $\theta \in (0, \infty)$.

Berger [1] proved a source coding theorem for the special case of a nonstationary autoregressive process with $a_1 = 1$ and $a_k = 0$ for $k > 1$ and he showed that the Kolmogorov formula still provided the rate-distortion function in this case. Gray [2] subsequently proved a source coding theorem for the general case described above and derived a rate-distortion function for this case resembling the Kolmogorov formula, but with (4) replaced by Eq. (22b) from [2]:

$$
R(D_\theta) = \frac{1}{2\pi}\int_{-\pi}^{\pi} \frac{1}{2}\ln\left(\max\left[g(\omega),\ \frac{\sigma^2}{\theta}\right]\right) d\omega. \qquad (6)
$$

Note that while (6) resembles the Kolmogorov formula, it is not the same. Both formulas derive from the finite dimensional versions of the Kolmogorov formula, that is, the finite order rate-distortion functions. But the mechanics of taking the limit from the finite order results differ in a critical way in that the integrand is bound away from zero, as will be described in more detail later. The equivalence of the two formulas follows in the stationary case because of the existence of source coding theorems for each, but it does not follow in the nonstationary case.

Let $R_K$ denote the Kolmogorov formula of (4) for stationary Gaussian processes applied to the autoregressive case, and let $R_{\text{AR}}$ denote the autoregressive formula of (6) and define the subset $E = \{\omega : g(\omega) < \sigma^2/\theta\}$ of $[-\pi, \pi]$. Then

$$
\begin{aligned}
R_{\text{AR}} - R_{\text{K}} &= \frac{1}{2\pi}\int_{-\pi}^{\pi} \frac{1}{2}\ln\left(\max\left[g(\omega), \frac{\sigma^2}{\theta}\right]\right) d\omega - \frac{1}{2\pi}\int_{-\pi}^{\pi} \max\left[\frac{1}{2}\ln\frac{\sigma^2}{\theta g(\omega)}, 0\right]\\
&= \frac{1}{2\pi}\int_{E^c}\left(\frac{1}{2}\ln g(\omega)\right) d\omega + \frac{\sigma^2}{\theta}\frac{1}{2\pi}\int_E d\omega - \frac{1}{2\pi}\int_E \left[\frac{1}{2}\ln\left(\frac{\sigma^2}{\theta g(\omega)}\right)\right]\\
&= \frac{1}{2}\frac{1}{2\pi}\int_{-\pi}^{\pi} \ln g(\omega)\, d\omega && (7)
\end{aligned}
$$

and hence the two formulas will not agree unless the final integral is 0.

In 1980 Hashimoto and Arimoto [3] revisited the question of the rate-distortion in the nonstationary case. They considered the finite order autoregressive case and noted that both the source coding-theorem and the evaluation of the rate-distortion function had been accomplished for the Wiener process in [1], but they only described the source coding theorem and not the rate distortion function of [2] for the more general autoregressive case, stating that "the rate-distortion function has not been calculated for nonstationary processes except for the Wiener process" and presented an "example which shows the form (3) is incorrect if the process is not asymptotically stationary, and we present the exact form of the rate-distortion in the next section." Their equation (3), however, corresponds to the Kolmogorov form of (4) and not the autoregressive form of (6), so that their example provided a demonstration that the Kolmogorov formula fails in the nonstationary case, but not that there was a problem with the autoregressive result (6) of [2]. As a result, there has been some confusion about the validity of the rate-distortion function of [2] in the nonstationary case and the apparently different result provided in [3] as well as some confusion about applicability of the specific asymptotic eigenvalue results for Toeplitz matrices used in [2].

We here reconcile the two forms for the nonstationary case and demonstrate that they are indeed consistent and distinct from the Kolmogorov formula in the nonstationary case. We also remark on some related issues regarding the eigenvalue distributions of certain asymptotically Toeplitz matrices.

## 2 Nonstationary autoregressive processes revisited

For the $M$th-order autoregressive process ($a_k = 0$ for $k > M$), Hashimoto and Arimoto correctly point out that the Kolmogorov formula (4) (their (3)) fails for a simple first order nonstationary autoregressive source and they state their main result, which replaces by the formula

$$R(D_\theta) = R_{\rm K} + \sum_{k=1}^{M} \max\left[\frac{1}{2}\ln|\rho_k|^2,\ 0\right] \tag{8}$$

where $\rho_k$ are the zeros of the characteristic polynomial

$$A(z) = \sum_{k=0}^{M} a_k z^{-k}. \tag{9}$$

Suppose that $|\rho_1| \geq | \cdots\ |\rho_m| > 1 > |\rho_{m+1}| \cdots\ \geq |\rho_M|$. Then, this can be rewritten as

$$R(D_\theta) = R_{\rm K} + \sum_{k=1}^{m} \frac{1}{2}\ln|\alpha_k|^2 \tag{10}$$

where $\alpha_k$ are the roots of $A(z)$ outside the unit circle.

On the other hand, the Jacobi-Jensen formula for analytic functions (e.g., [6], p. 23, or [7], p. 207) applied to (7) yields

$$R_{\rm AR} = R_{\rm K} + \sum_{k=1}^{m} \frac{1}{2}\ln|\alpha_k|^2, \tag{11}$$

which agrees with the rate-distortion function of (10). Note that since $g(\omega)$ is analytic, it can have at most a finite number of zeros outside the unit circle. Thus, in particular, the results of [3] demonstrate that the Kolmogorov formula may fail for nonstationary sources, not that the autoregressive formula is incorrect. The two formulas agree for stationary sources and for the nonstationary Wiener process.

## 3 Asymptotic eigenvalue distributions

Although the rate-distortion functions of [2] and [3] are equivalent, they use different versions of the classic asymptotic eigenvalue distribution theorem for Toeplitz matrices. The classic form can be described as follows. Given a discrete-time Fourier transform pair

$$f(\omega) = \sum_{k=-\infty}^{\infty} t_k e^{-jk\omega} \tag{12}$$

$$t_k = \frac{1}{2\pi}\int_{-\pi}^{\pi} f(\omega) e^{jk\omega}\, d\omega \tag{13}$$

let $T_n = \{t_{k-\ell}; k, \ell = 0, 1, \ldots, n-1\}$ be the corresponding Toeplitz matrix with eigenvalues $\tau_{n,k}$; $k = 0, 1, \ldots, n-1$. Suppose that the essential infimum and supremum of $f$ by $m_f$ and $M_f$, respectively. Then the classical theorem states that if $F$ is a continuous function on $[m_f, M_f]$

$$\lim_{n\to\infty} \frac{1}{n}\sum_{k=0}^{n-1} F(\tau_{n,k}) = \frac{1}{2\pi}\int_{-\pi}^{\pi} F(f(\omega))\, d\omega. \tag{14}$$

If any sequence of matrices $B_n$ is asymptotically equivalent to $T_n$ in the sense of being bounded and having a vanishing Hilbert-Schmidt norm $|B_n - T_n|$, then (14) will also hold for its eigenvalues.

The classic Kolmogorov result for stationary autoregressive processes follows from his finite order results by taking $T_n$ as the $n$th order covariance matrix of the Gaussian process, $t_{k-j} = K_X(k, j)$, and using the Toeplitz limit to compute

$$D_\theta = \lim_{n\to\infty} \frac{1}{n}\sum_{k=0}^{n-1} \min(\theta,\ \tau_{n,k}) \tag{15}$$

$$R(D_\theta) = \lim_{n\to\infty} \frac{1}{n}\sum_{k=0}^{n-1} \max\left(0,\ \frac{1}{2}\ln\frac{\tau_{n,k}}{\theta}\right). \tag{16}$$

The autoregressive result, however, instead focuses on the inverse covariance. The difference equation defining an autoregressive process can be written in vector form as

$$A_n X^n = Z^n.$$

where the lower triangular Toeplitz matrix $A_n$ is given by

$$A_n = \begin{bmatrix} 1 & & & & \\ a_1 & 1 & & & 0 \\ & a_1 & 1 & & \\ & & \ddots & \ddots & \\ & & & & \\ a_{n-1} & & & a_1 & 1 \end{bmatrix}. \tag{17}$$

The inverse covariance

$$(K_X^{(n)})^{-1} = \sigma^{-2} A_n^* A_n. \tag{18}$$

is then asymptotically equivalent to the Toeplitz matrix $T_n(g(\omega)/\sigma^2)$, where $g(\omega)$ is given by (5) and hence the Toeplitz eigenvalue distribution theorem can be applied with $\tau_{n,k} = 1/\lambda_{n.k}$, where the $\lambda_{n,k}$ are the eigenvalues of $\sigma^{-2} A_n^* A_n$.

As Hashimoto and Arimoto point out, in the nonstationary case direct application of the asymptotic eigenvalue distribution theorem does not work in evaluating the limit of (16) because of the behavior of the $\lambda_{n,k}$ near zero. Alternatively, $\ln r$ is not continuous at $r = 0$ and hence the conditions of the eigenvalue distribution theorem do not apply. This difficulty is obvious from rewriting (16) as

$$R(D_\theta) = \lim_{n\to\infty} \frac{1}{n} \sum_{k=0}^{n-1} \max\left(0,\ \frac{1}{2}\ln\frac{1}{\lambda_{n,k}\theta}\right)$$

since the $\lambda_{n,k}$ are not bound away from 0. The observation in [3] is that exactly the $m$ smallest $\lambda_{n,k}$ decrease exponentially as $n$ increases while the remaining $\lambda_{n,k}$ are bounded from zero. Between those $m$ smallest $\lambda_{n,k}$, the $\ell$th smallest one decreases asymptotically as $|\rho_\ell|^{-2n}$, for $\ell = 1, 2, \cdots, m$, and the expression (8) follows.

The derivation of [2], however, avoided the above difficulty by deriving an equivalent form to the Kolmogorov finite order formula:

$$\frac{1}{n} \sum_{k=0}^{n-1} \max\left(0,\ \frac{1}{2}\ln\frac{1}{\lambda_{n,k}\theta}\right) = \frac{1}{n} \sum_{k=0}^{n-1} \ln\left[\max\left(\frac{\sigma^2}{\lambda_{n,k}},\ \frac{\sigma^2}{\theta}\right)\right] \tag{19}$$

and then applying a variation of the Toeplitz eigenvalue theorem for functions that are confined to the region where the eigenvalues are bounded away from 0, that is, the Toeplitz theorem is applied not to the function $F(\lambda) = \max(0, \frac{1}{2}\ln 1/\lambda\theta)$ as in the classic Kolmogorov case, but to the function $F(\lambda) = \max\left[\sigma^2/\lambda, \sigma^2/\theta\right]$ (see the discussion between (21) and (22) in [2]). This yields an answer with a different functional form which is not contradicted by the example of [3] and which has no problems with $\lambda_{n,k}$ near 0.

This trick of truncating both the sum and integral to avoid problems at singularities has been well developed in the literature and the general result is mentioned below as an indication of how singularities such as arise in nonstationary autoregressive processes are easily handled in the asymptotic eigenvalue distribution theory.

In some applications we wish to study the asymptotic distribution of a function $F(\tau_{n,k})$ of the eigenvalues of an asymptotically Toeplitz sequence of matrices that is not continuous at the minimum or maximum value of $f$. For example, in order for results derived to apply to the function $F(f(\lambda)) = 1/f(\lambda)$ which arises when treating inverses of Toeplitz matrices, it is often considered necessary to require that the essential infimum $m_f > 0$ because the function $F(1/x)$ is not continuous at $x = 0$. If $m_f = 0$, the basic asymptotic eigenvalue distribution breaks down and the limits and the integrals involved might not exist — the limits might exist and equal something else or they might simply fail to exist.

In order to treat the inverses of Toeplitz matrices when $f$ has zeros, define the mid function

$$\mathrm{mid}(x, y, z) \triangleq \begin{cases} z, & y \geq z \\ y, & x \leq y \leq z \\ x, & y \leq z \end{cases} \tag{20}$$

The following result was proved in [9] and extended in [10]. See also [11, 12, 13, 8].

**Theorem 1** *Suppose that $f$ is in the Wiener class. Then for any function $F(x)$ continuous on $[\psi, \theta] \subset [m_f, M_f]$*

$$\lim_{n\to\infty} \frac{1}{n} \sum_{k=0}^{n-1} F(\mathrm{mid}(\psi,\ \tau_{n,k},\ \theta)) = \frac{1}{2\pi} \int_0^{2\pi} F(\mathrm{mid}(\psi,\ f(\lambda),\ \theta)\, d\lambda. \tag{21}$$

This asymptotic eigenvalue distribution theorem yields the rate-distortion function in the nonstationary autoregressive case and avoids the singularity problems encountered by [3]. It is stated in [3] that the asymptotic eigenvalue distributions of $T_n(g)$ and $\sigma^{-2}A_n^*A_n$ are *not* the same and this fact is demonstrated by the example of a first order nonstationary autoregressive Gaussian process which violates (4), but this is only true in the strict sense that the traditional eigenvalue distribution theorem does not hold for the function $F$ considered. These two matrix sequences, however, are asymptotically equivalent and their eigenvalue distributions *do* satisfy the truncated form of Theorem 1. Thus the eigenvalues are indeed asymptotically equally distributed, provided they are cut off at suitable values.

In conclusion, the results of [2] and [3] are consistent and the results of the latter provide no evidence of invalidity of the former. The two papers provide alternative characterizations of the same quantity which are related through the Jacobi-Jensen formula. The second paper provided the first detailed example where the Kolmogorov and autoregressive formulas for the rate-distortion function differed by a nonzero amount.